\RequirePackage{fix-cm}
\documentclass[smallextended]{svjour3}       
\smartqed  
\usepackage{graphicx}
\usepackage{listings}
\usepackage{color}
\usepackage{url}
\usepackage{array}
\usepackage{multirow}
\usepackage{amsmath}
\usepackage{multirow}
\usepackage{booktabs}
\usepackage{times}
\definecolor{DarkRed}{rgb}{.6,0,0}
\definecolor{DarkGreen}{rgb}{0,.6,0}
\definecolor{DarkBlue}{rgb}{0,0,.6}
\definecolor{Gray}{rgb}{.3,.3,.3}
\newcommand\addfigure[4]{
  \begin{#1}[loc=htbp] 
    \center \includegraphics[width=#2\linewidth]{#3}
    \caption{#4}\label{fig:#3}
  \end{#1}
}

\begin{document}

\title{soCloud: A service-oriented component-based PaaS for managing portability, provisioning, elasticity, and high availability across multiple clouds
}


\author{Fawaz Paraiso         \and
        Philippe Merle   	  \and
        Lionel Seinturier
}


\institute{Fawaz Paraiso \at
              Inria Lille - Nord Europe \&  University Lille 1\\
              LIFL UMR CNRS 8022, France\\
              \email{fawaz.paraiso@inria.fr}          
           \and
           Philippe Merle \at
              Inria Lille - Nord Europe \&  University Lille 1\\
              LIFL UMR CNRS 8022, France\\
              \email{philippe.merle@inria.fr}         
           \and
           Lionel Seinturier \at
              Inria Lille - Nord Europe \&  University Lille 1\\
              LIFL UMR CNRS 8022, France\\
              \emph{IUF - Institut Universitaire de France} \\
              \email{lionel.seinturier@inria.fr}                    		
}

\date{Received: 12 th July, 2013 / Accepted: date}

\maketitle

\begin{abstract}
Multi-cloud computing is a promising paradigm to support very large scale world wide distributed applications. Multi-cloud computing is the usage of multiple, independent cloud environments, which assumed no priori agreement between cloud providers or third party. However, multi-cloud computing has to face several key challenges such as \textit{portability}, \textit{provisioning}, \textit{elasticity}, and \textit{high availability}. Developers will not only have to deploy applications to a specific cloud, but will also have to consider application portability from one cloud to another, and to deploy distributed applications spanning multiple clouds. This article presents soCloud a service-oriented component-based Platform as a Service (PaaS) for managing portability, elasticity, provisioning, and high availability across multiple clouds. soCloud is based on the OASIS Service Component Architecture (SCA) standard in order to address portability. soCloud provides services for managing provisioning, elasticity, and high availability across multiple clouds. soCloud has been deployed and evaluated on top of ten existing cloud providers: Windows Azure, DELL KACE, Amazon EC2, CloudBees, OpenShift, dotCloud, Jelastic, Heroku, Appfog, and an Eucalyptus private cloud.

\end{abstract}

\section{Introduction}\label{sec:intro}
Cloud computing builds on established trends for driving the cost out of the delivery of services while increasing the speed and agility with which services are deployed. Virtualization, on-demand deployment, Internet delivery of services are parts of Cloud computing. Cloud computing differentiates itself by changing how we invent, develop, deploy, scale, update, maintain, and pay for applications and the infrastructure on which they run.

Different cloud service providers, based on different technologies, support a large number of cloud services such as Infrastructure as a Service (IaaS) and Platform as a Service (PaaS). Cloud service consumers select what fit their requirements from the cloud services. For instance, requirements can be: price, quality of service (QoS), programming language, database, middleware, etc. It is difficult to cloud service consumers to meet all these requirements with a single cloud provider. Multi-cloud computing as the usage of multiple, independent cloud environments, which assumed no priori agreement between cloud providers or third party is a promising paradigm to support very large scale world wide distributed applications.

However, multi-cloud computing has to face several key challenges: \textit{portability}, \textit{provisioning}, \textit{elasticity}, and \textit{high availability}. Multi-cloud portability means writing applications once and running them on any clouds. Most existing cloud providers are typically offered through proprietary APIs and limited to a single infrastructure provider. In such situations, vendor lock-in is a primary concern for moving towards a cloud provider. Multi-cloud provisioning refers to the capability to deploy a distributed application spanning multiple cloud providers. Deploying a distributed application in a multi-cloud context is not an easy task. Multi-cloud elasticity refers to the capability to scale applications across multiple clouds. Currently, there is no convenient way to express specific application elasticity rules for each part of a distributed application as needed. Multi-cloud high availability refers to the degree to which an application is operable across multiple clouds. Cloud provider services can become unavailable due to outages or denials of services. High availability needs to be analysed and set across multiple clouds in order to reduce the probability of outages that could affect services deployed in a single cloud system.

In this article we discuss the design and implementation of soCloud. soCloud is a multi-cloud PaaS that addresses the four key challenges presented previously. soCloud is a distributed PaaS that provides a model for building distributed applications. This model is an extension of the OASIS SCA standard\footnote{http://www.oasis-opencsa.org/sca}. Our ongoing approach to address portability and provisioning in a multi-cloud context is the use of the SCA standard. Our elasticity management approach is based on autonomic computing with the overall aim of creating self-managed elastic multi-cloud applications. High availability is achieved in two ways. Firstly, soCloud provides a multi-cloud load balancer service that fronts traffic for applications deployed across multiple clouds and makes a decision about where to route the traffic when cloud nodes fail. Secondly, the soCloud architecture uses redundancy at all levels to ensure that no single component failure in a cloud provider impacts the overall system availability. We describe a way to annotate SCA artifacts with deployment information needed to optimize the use of services in multiple cloud environments. These annotations also allow to express elasticity rules that ensure the appropriate adjustment decisions made in timely manner to meet service needs in the presence of cloud service failures. The soCloud architecture is composed of the following SCA components: service deployer, constraints validator, PaaS deployment, SaaS deployment, load balancer, node provisioning, monitoring, workload manager and controller components. soCloud is deployed and evaluated on ten existing cloud providers Windows Azure, DELL KACE, Amazon EC2, CloudBees, OpenShift, dotCloud, Jelastic, Heroku, Appfog, and an Eucalyptus private cloud.

The remainder of this article is organized as follows. In Section~\ref{sec:chal}, we discuss the four challenges we addressed for multi-clouds. Next, Section~\ref{sec:over} presents the design and implementation of the soCloud platform, and its integration with existing cloud providers. The evaluation of soCloud is discussed in Section~\ref{sec:valid}. Section~\ref{sec:related} compares soCloud with the state-of-the-art. Section~\ref{sec:discus} discusses the limitations of this work, while Section~\ref{sec:conclusion} concludes this article and presents future work we intend to address.

\vspace{-0.5cm}
\section{Multi-cloud challenges}\label{sec:chal}
IT companies are starting to realize and recognize the benefits and advantages of cloud computing. However, cloud technology maturity is still a concern. This section describes four key challenges for  multi-cloud computing: \textit{portability}, \textit{provisioning}, \textit{elasticity}, and \textit{high availability}.
\vspace{-0.5cm}
\subsection{Multi-cloud portability}\label{subsec:porta}
\vspace{-0.2cm}
In the cloud computing area, the portability issue should take into account both {\it application} and {\it data}. Although data portability is an important feature, this article focuses only on application portability.
In an emerging and rapidly changing market such as cloud computing, it is easy to create applications that are locked into one vendor cloud because of the use of proprietary APIs and formats. To avoid this vendor lock-in syndrome, SaaS must be portable on top of various cloud PaaS and IaaS providers. Then, this multi-cloud portability allows the migration from one cloud provider to another in order to take advantage of cheaper  prices or better QoS. However, SaaS portability requires that the runtime support provides a common model to hide the diversity of underlying PaaS and IaaS. Furthermore, the dominant programming models today have grown increasingly complex. SCA, in contrast, provides a simplified programming model and unified way to applications that communicate using a variety of network protocols~\cite{marino2010understanding}. 

To address the challenge of \textit{multi-cloud portability}, soCloud promotes SCA as the model to design and develop both multi-cloud SaaS applications and the underlying soCloud PaaS.
\vspace{-0.5cm}
\subsection{Multi-cloud provisioning}
\vspace{-0.2cm}
\paragraph{\textbf{Application Provisioning:}} Application provisioning inclu\-des building and deployment on multiple cloud environments. Providing a consistent methodology and process for modelling how applications are built and provisioned, enabling flexibility and choice for developers to use any cloud provider they choose. Application provisioning should deliver business agility and operational efficient by high level abstraction and automating provisioning of applications across multiple cloud providers.
\vspace{-0.3cm}
\paragraph{\textbf{Geo-diversity:}} The authors in~\cite{zhang2010cloud} advocates that small data centers, which consume less power, may be more advantageous than large ones, and that geo-diversity tends to better match user demands. Geo-diversity lowers latency to users and increases reliability in the presence of an outage taken out an entire site. In a legal context, data protection law and confidentiality can lead users to place their data in a specific area. In fact, the location of data can be facilitated or restricted in particular jurisdictions.

Overall, to address the challenge of \textit{multi-cloud provisioning}, soCloud offers a service to provision applications across multiple cloud providers.
\vspace{-0.5cm}
\subsection{Multi-cloud elasticity}
\vspace{-0.2cm}
The management of elasticity can be further split into two approaches: fine-grained or coarse-grained. The first one allows to scale resources either by changing the number of virtual machines (VMs) using horizontal scaling (adding more virtual machines or devices to the computing platform to handle an increased application load) or vertical scaling (adding more CPU, Memory, Disk, Bandwidth to handle an increased application load) depending on the application memory, storage, network bandwidth and CPU requirements. The second one manages the resources scalability by changing cloud providers. Indeed, when outages occur with one cloud provider, the coarse-grained elasticity will switch to another cloud provider. While, the fine-grained elasticity can actually be made up of many fine-grained resources. In managing elasticity across multiple clouds, automation is a mandatory requirement, and it is thus a foundational design principle~\cite{isard2007autopilot}. The function of any autonomic capability is a control loop that collects details from the system and acts accordingly. However, developers should have the possibility to define specific elasticity rules on their services. For example, the developers specify constraints on the response time depending of the number of users currently accessing the provided service. 

To address the challenge of \textit{multi-cloud elasticity}, soCloud offers an autonomic service which provides a global mechanism to manage elasticity across multiple clouds and also offers the possibility to define application specific elasticity rules.
\vspace{-0.5cm}
\subsection{Multi-cloud high availability}
\vspace{-0.2cm}
A series of news \cite{news1,news2} and papers~\cite{armbrust2010view,paraiso:hal-00790455} have pointed several cloud provider outages. According to a recent report by the International Working Group on Cloud computing Resiliency\footnote{http://iwgcr.org}, a total of 568 hours of downtime at thirteen well-known cloud services since 2007 caused financial damage of more than US\$71.7 million. The average unavailability of cloud services is 7.5 hours per year, amounting to an availability rate of 99.9\%, according to the group preliminary results. These results are far from the expected reliability of mission critical system which is 99.999\%. As a comparison, the average unavailability for electricity in a modern capital city is less than 15 minutes per year~\cite{iwgcr}. Besides this economic impact, the downtime also affects millions of end-users. Of course, downtime costs money and dammage, unfortunately protecting systems against downtime with 99.999\% of availability is not free. 

To address the challenge of \textit{multi-cloud high availability} despite outages, soCloud provides high availability in two ways. Firstly, with the applications deployed with a soCloud platform, the high availability is ensured by using a load balancer service which distributes requests among instances of the application deployed on multiple cloud providers. Secondly, the soCloud architecture uses redundancy at all levels to ensure that no single component failure in a cloud provider impacts the overall system availability.

\vspace{-0.5cm}
\section{soCloud design and implementation}\label{sec:over}
In this section we present the design and implementation of soCloud platform. We first discuss background elements of SCA and FraSCAti. Next, we describe some components of the soCloud architecture and its implementation. Finally, we describe how the soCloud platform is deployed on existing IaaS/PaaS providers.
\vspace{-0.5cm}
\subsection{SCA}
\vspace{-0.2cm}
soCloud is based on the SCA standard. SCA is a set of OASIS specifications for building distributed applications and systems using Service-Oriented Architecture (SOA) principles~\cite{erl2008soa}. SCA promotes a vision of Service-Oriented Computing (SOC) where services are independent of implementation languages (Java, Spring, BPEL, C++, COBOL, C, etc.), networked service access technologies (Web Services, JMS, etc.), interface definition languages (WSDL, Java, etc.) and non-functional properties. Component-Based Design~\cite{chen2007refinement} and SOA are two major software engineering approaches widely used for structuring systems. SCA targets composition of services in SOA systems and thus is suitable for building enterprise and cross-enterprise applications built on already-developed components and services. 
\vspace{-0.5cm}
\subsection{FraSCAti}
\vspace{-0.2cm}
Several open source implementations of the SCA specifications exist. Three of the most well known are Apache Tuscany, Fabric3 and FraSCAti. Compared to Tuscany and Fabric3, FraSCAti introduces reflective capabilities to the SCA programming model, and allows dynamic introspection and reconfiguration via a specialization of the Fractal component model~\cite{Fractal}. FraSCAti provides a component-based approach to support the heterogeneous composition of various interface definition languages (WSDL, Java), implementation technologies (Spring, EJB, BPEL, OSGI, Jython, Jruby, Xquery, Groovy, Velocity, Fscript, Beanshell.), and binding technologies (Web Services, JMS, RPC, REST, RMI, UPnP.).

soCloud is built on top of FraSCAti. FraSCAti is the execution environment of both the soCloud PaaS and soCloud applications deployed on the top of this multi-cloud PaaS.
\vspace{-0.5cm}
\subsection{soCloud SaaS applications}\label{subsec:sapp}
\begin{figure*}[htbp]
\begin{center}
\centering
\includegraphics[height=0.5\textwidth, width=1.0\linewidth]{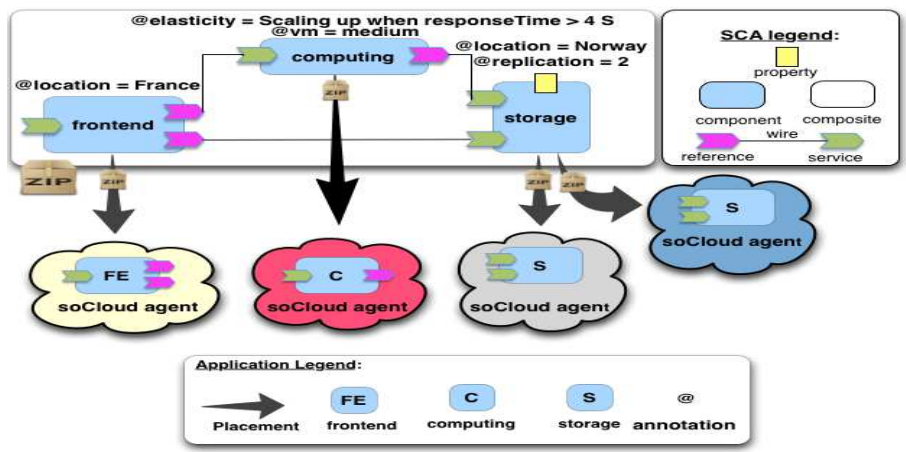}
\caption{An annotated soCloud application.}
\label{fig:three-tier-architecture}
\end{center}
\end{figure*}

\vspace{-0.2cm}
\paragraph{\textbf{Application specification}} soCloud applications are built using the SCA model. As illustrated in Fig.~\ref{fig:three-tier-architecture}, the basic SCA building blocks are software components, which provide services, require references and expose properties. The references and services are connected by wires. For SCA references, a binding describes the access mechanism used to invoke a remote service. In the case of services, a binding describes the access mechanism that clients use to invoke the service. We describe how SCA can be used to package SaaS applications. The first requirement is that the package must describe and contain all artifacts needed for the application. The second requirement is that provisioning constraints and elasticity rules must be described in the package. The SCA assembly model specification describes how SCA and non-SCA artifacts (such as code files) are packaged. The central unit of deployment in SCA is a contribution. A contribution is a package that contains implementations, interfaces and other artifacts necessary to run components. The SCA packaging format is based on ZIP files, however, other packaging formats are explicitly allowed. Fig.~\ref{fig:three-tier-architecture} shows a three-tier application is packaged as a ZIP file (SCA contribution) and its architecture is described. 
\vspace{-0.5cm}
\paragraph{\textbf{Annotations}} Some cloud-based applications require more detailed description of their deployment (c.f. Fig.~\ref{fig:three-tier-architecture}). The deployment and monitoring of soCloud applications are bound by a description of the overall software system architecture and the requirements of the underlying components, that we refer to as the \emph{application manifest}. Basically, the \emph{application manifest} consists of describing what components the application is composed with functional and non-functional requirements for deployment. In fact, the application can be composed of multiple components (c.f. Fig.~\ref{fig:three-tier-architecture}). The application manifest defines elasticity rule for the service component (e.g., increase/decrease instance of component). Commonly, scale up or down, is translated to a \textbf{condition-action} statement that reasons on performance indicators of the component deployed. In order to fulfill the requirements for the soCloud application descriptor, we propose to annotate the SCA components with the four following annotations:
\begin{enumerate}
  \item \textbf{placement constraint} (\textit{@location}) allows to map components of a soCloud application to available physical hosts within a geographical datacenter in multi-cloud environments.
  \item \textbf{computing constraint} (\textit{@vm}) provides necessary computing resources defined for components of a soCloud application in the multi-cloud environments.
  \item \textbf{replication} (\textit{@replication}) specifies the number of instances of the component that must be deployed in multi-cloud environments.
  \item \textbf{elasticity rule} (\textit{@elasticity}) defines a specific elasticity rule that should be applied to the component deployed on multi-cloud environments.
\end{enumerate}

For example, let us consider the three-tier web application described in Fig.~\ref{fig:three-tier-architecture}. The annotation (\textit{@location=France}) of the frontend component indicates to deploy this component on a cloud provider located in France. Next, the annotation (\textit{@vm=medium}) on the computing component specifies the kind of computing resources required by this component and can be deployed on any cloud provider. The developer has the possibility to specify through the \textit{@vm} annotation the computing resources (micro, small, medium, large) she need. Finally, the annotations (\textit{@location=Norway} and \textit{@replication=2}) on the storage component indicate to deploy this component on two different cloud providers located in Norway. soCloud automates the deployment of this three-tier application in a multiple cloud environment by respecting given annotations.
\vspace{-0.5cm}
\subsection{Constraint analysis and formulation}
\vspace{-0.2cm}
To express constraints (placement, computation, etc.) and define specific elasticity rules, we analyse each step of the formulation of these constraints. To express a constraint we use this formula P = \{n, v\}. Where \textit{n} indicates the name of the constraint and \textit{v} the value of the constraint. Regarding the elasticity rule, we use R = \{c, a\}, where \textit{c} indicates the condition and \textit{a} the resulting action.

The placement can be a location or a provider name. For example, specifying {\it @placement="Amazon\_Ireland"} or {@placement="Ireland"} on a component has the same interpretation (i.e, the component should be placed in Ireland) from the point of view of {\it placement constraint}.

To express different computation capacities, we use the instance type taxonomy defined by Amazon EC2\footnote{http://aws.amazon.com/ec2/instance-types/}. An example of computation constraint request is P = \{vm, medium\}. This request has \textit{name} ``vm'' and value ``medium'' which represents the computing capacity.

However, in a heterogeneous multi-cloud environment, existing types of VM provided by different clouds can have small differences. In order to use these clouds and hide these differences, the soCloud platform defines a high-level abstraction where similar VMs are classified in the same type.
Let \( T\) be the set of VM types defined in the soCloud platform (see Equation~\ref{eq:type}) and \( C\) the characteristics of the VM provided by different clouds (see Equation~\ref{eq:car}). Equation~\ref{eq:vmtype} defines how the soCloud platform hides the differences between the VMs provided from different clouds.

\begin{equation}
T = \{ micro, small, medium, large \}
\label{eq:type}
\end{equation}
\vspace{-5mm}
\begin{equation}
C = \{ C_{i}, i \in Provider \}
\label{eq:car}
\end{equation}
\vspace{-5mm}
\begin{equation}
VM_{type} = \{ C : f(C_{i}), f(C_{i}) \in T, i \in Provider \}
\label{eq:vmtype}
\end{equation}

The soCloud platform offers to developers the opportunity to choose the specific VM (with the best performance) by indicating, as an additional information, {\it the provider name}. 
Overall, choosing a specific VM refers to the combination of {\it @vm} and {@placement} annotations, where {@placement} corresponds to the provider name.

\vspace{-0.5cm}
\subsection{A soCloud application descriptor}
\vspace{-0.2cm}
Let us consider the three-tier application architecture described in Fig.~\ref{fig:three-tier-architecture}, where we need to deploy a distributed application. This distributed application is packaged as a contribution that contains three contributions and one file (\textit{application descriptor}) describing the architecture of the distributed application. Each contained contribution corresponds to each tier of the distributed application. In the case the application deployed is not distributed, the contribution contains a single contribution with a composite file. This distributed application needs placement, elasticity, and computation requirements. In order to fulfill these requirements, we use SCA properties to express them (see Listing~\ref{lst:sca}). Lines 3, 9, 18 correspond to our SCA extension defined to represent respectively frontend, computing and storage contributions. The placement constraints for frontend and storage components are expressed at Line 6 and 20 respectively. The computing constraint is expressed at Line 12 for the computing component. The number of replication of the storage component is expressed at Lines 21. Lines 13-15 express the elasticity rule for the computing component. We adopt an event-condition-action approach for rule specification. The event-condition syntax is an Event Processing Language statement~\cite{cep}. Basically, the elasticity rule and action defined at Lines 13-15 means: when the average response time of the component exceeds 4 seconds, then add a new virtual machine running this component. 

\begin{minipage}{1.0\textwidth}
\begin{scriptsize}
\begin{lstlisting}[emph={interface.java,implementation.java},numbers=left,caption={A soCloud application descriptor.},label={lst:sca}]
<composite name="DistributedApplication">
  <component name="frontend">
    <implementation.contribution contribution="frontend.zip"/>
    <reference name="compute" target="computing/compute"/>
    <reference name="storage" target="storage/storage"/>
    <property name="location">France</property>
  </component>
  <component name="computing">
    <implementation.contribution contribution="computing.zip"/>
    <service name="compute"/>
    <reference name="storage" target="storage/storage"/>
    <property name="vm">medium</property>
    <property name="elasticity">
    Scaling out when ResponseTime > 4s
    </property>
  </component>
  <component name="storage">
    <implementation.contribution contribution="storage.zip"/>
    <service name="storage"/>
    <property name="location">Norway</property>
    <property name="replication">2</property>
  </component>
</composite> 
\end{lstlisting}
\end{scriptsize}
\end{minipage}

\vspace{-0.7cm}
\subsection{soCloud architecture}
\begin{figure*}[htbp]
\begin{center}
\centering
\includegraphics[height=0.45\textwidth, width=1.0\linewidth]{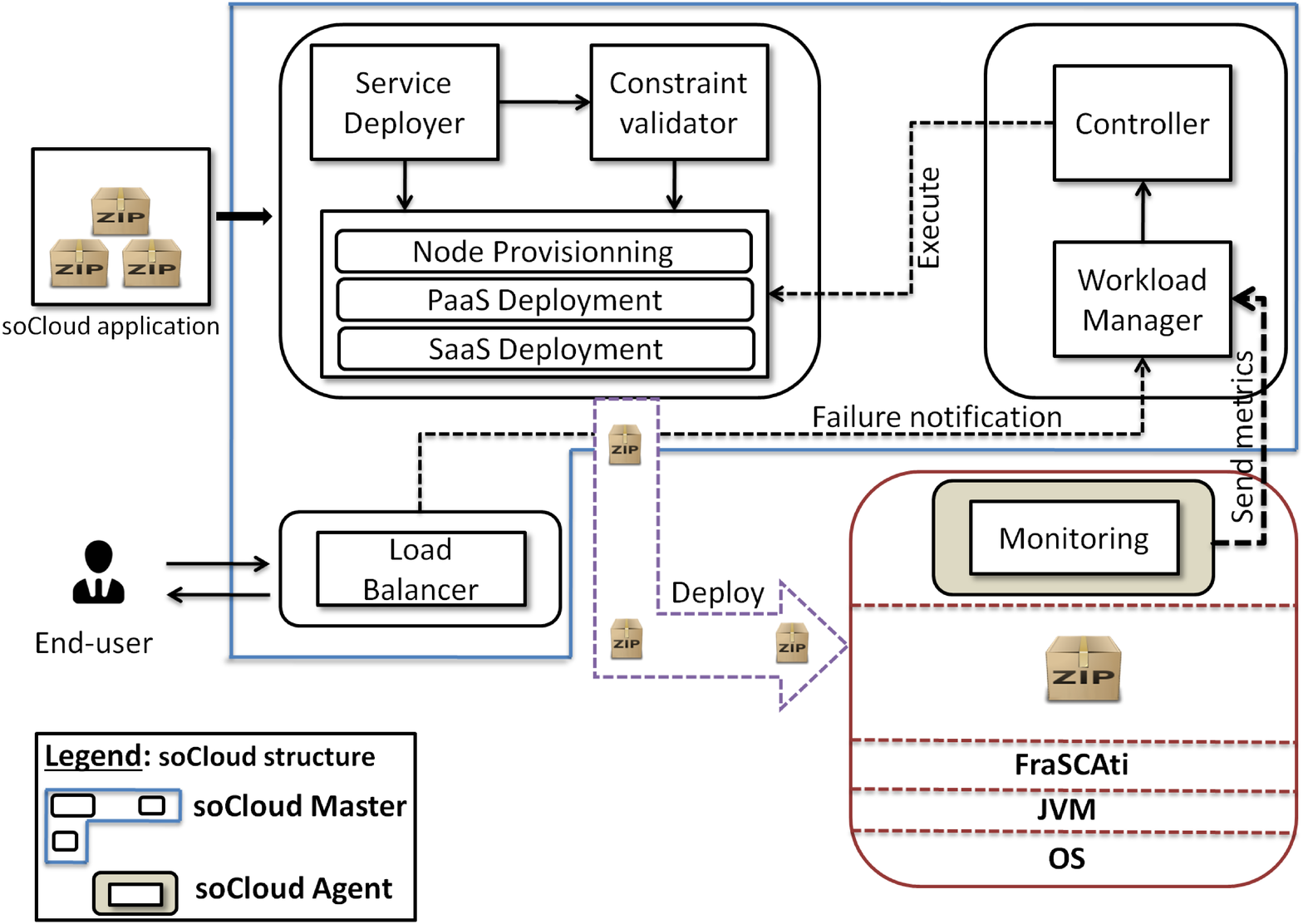}
\caption{Overview of the soCloud Architecture.}
\label{fig:mcp}
\end{center}
\end{figure*}

\vspace{-0.2cm}
Fig.~\ref{fig:mcp} gives an overview of the soCloud architecture. The soCloud architecture has two parts: the soCloud \emph{master}, and the soCloud \emph{agent}. This partitioning provides flexibility for deploying the soCloud PaaS across a highly distributed multi-cloud environment. Firstly, the soCloud \emph{master} consists of a set of eight components. This part of the architecture focuses on the intelligence processing of soCloud. Secondly, the soCloud \emph{agent} is used to host, execute and monitor soCloud applications. This part provides the necessary services for managing a set of applications and resources. soCloud \emph{agents} work with the soCloud \emph{master} and run in different cloud infrastructures. All communication between a soCloud \emph{master} and the applications deployed is mediated by the soCloud \emph{agent}. 

\vspace{-0.5cm}
\subsubsection{Monitoring}
\vspace{-0.2cm}
The component provides an unified-platform independent mechanism that collects, aggregates and reports details (such as health and performance metrics) about applications deployed on multiple cloud environments. It brings information about currently executing process as well as the system on which the monitoring service is running. The \emph{monitoring} component captures any change in the state of the application. The monitoring associates to each application deployed on the soCloud platform, a temporary table (ResponsivenessEvent) that collects informations such as application \textit{responseTime}, \textit{number of requests}, etc. The metrics collected in a time interval are sent to the \emph{Workload Manager} component for analyzing. The monitoring component acts at three levels: Operating System (OS), Java Virtual Machine (JVM), and Execution Environment (FraSCAti). The monitoring component exposes services via REST and JMS to monitor a distributed environment. The consumer of these services is the \emph{workload manager} component or can be also any external application running outside soCloud. Each application deployed with the soCloud PaaS is automatically monitored. However, with some cloud providers such as Salesforce.com or Google App Engine our monitoring component could not work. As example, Google App Engine forbids the use of JMX.

\vspace{-0.4cm}
\subsubsection{Workload Manager}
\vspace{-0.2cm}
The Workload Manager (WM) component provides some event processing functionality~\cite{eventprocessing}. All events are processed to extract drift indicators (DI). An example of DI can be a CPU consumption is greater than 90\% for a period of 2 minutes. The WM is centered on DI tracking perform filtering, transformation, and most importantly aggregation of events. All the metrics (events data) sent by \emph{monitoring} components are continuously analyzed in terms of drift indicators that are expressed by event rules, and acts upon opportunities and threats in real time, potentially by creating derived events. One of WM major goals is to find a symptom and analyses it to find its root cause. The WM uses a technique called event \emph{correlation}\footnote{http://tinyurl.com/qdrcpm3.} to examine symptoms and identify groups of symptoms that have a common root cause. As an example of event correlation, WM takes multiple occurrences of the same event, examines them for duplicate information, removes redundancies and reports them as a single event. When a drift occurs, the WM reports it to the \emph{controller} component. The ability to derive instant insights into the operations of the resource provisioning is essential. Thus, the capability to dynamically allocate and dispose resources is an important ingredient to build a platform for elastic applications. 

Related to the events received from an inbound \emph{monitoring} component, how events can be woven together to pull out the right information? This is accomplished through Complex Event Processing (CEP)\footnote{{\bf CEP}: Computing that performs operations on complex events, including reading, creating, transforming, or abstracting them\cite{cep}.}. To achieve this, we use DiCEPE, a Distributed Complex Event Processing Engine we have presented in~\cite{paraiso2012middleware}. The particularity of DiCEPE is the integration of CEP engines in distributed systems, and the fact that they can be exposed via various communication protocols. The DiCEPE integrates the Esper\footnote{http://www.espertech.com/} engine for further processing. We apply Esper because of its performance and the metric-value pairs are delivered as events each time their values change between measurements. 

\vspace{-0.4cm}
\subsubsection{Controller}
\vspace{-0.2cm}
The \emph{controller} component provides the mechanisms that construct the actions needed to achieve goals and objectives. For example, it multiplexes workloads onto an existing infrastructure, and allows for on-demand allocation of resources to workloads. The system state is managed by the \emph{controller} component. By state, we mean the information retained in one component that is meaningful for this component (as example: a table on each instance of LB to associate network addresses with the symbolic names of available hosts). The system state offers the potential for improving the consistency, and reliability of the system. For components to work together effectively, they must agree on common goals and coordinate their actions. This requires that each part to know something about the other. For example, the \emph{node provisioning} stores a table of available resources: If the developer wants to deploy an application on the resource, the controller can notify the \emph{node provisioning} to allocate new resources for the application when the available resource is not sufficient. The second potential advantage of the system state is reliability. If information is replicated at several cloud providers and one of the copies is lost due to a failure, then it may be possible to use one of the other copies to recover the lost information. Compared to the \emph{workload manager} and \emph{node provisioning} components, the \emph{controller} takes decision in the system. The \emph{controller} component is self-adaptive in order to respond in a coherent and timely manner to changes in environment, and to failures of components.

All requests handled by a \emph{controller} component are processed as transactions. The transaction engine is implemented for the specific needs of the soCloud architecture. Each transaction is created and managed by a coordinator. Two well-known problems of concurrent transactions can be mentioned: i) lost update and ii) inconsistent retrievals. To avoid these problems we use a serially equivalent execution of transactions~\cite{coulouris2005distributed}. The use of serial equivalence as a criterion for correct concurrent execution prevents the occurrence of lost updates and inconsistent retrievals. The \emph{controller} component is the core of the elasticity management, it is made to tolerate failures by the use of redundant components. 

\vspace{-0.4cm}
\subsubsection{Service Deployer}
\vspace{-0.2cm}
The process illustrated by the sequence diagram in Fig.~\ref{fig:sequence-diagram} describes how each task vary with a service deployment scenario.
The \emph{Service Deployer} (SD) component is responsible for handling the additional information of coordinating and managing the service across multiple clouds (i.e., placement, binding, manage service). The SD component decomposes and captures the constraints (specified by the developer) of the service. For example, a constraint can be a placement of an application, resource capacities needed by an application, or defined elasticity rules. In the case where the constraints expressed on the components are fulfilled by multiple cloud providers, soCloud randomly choses one of the providers offering the lowest price.
\addfigure{figure}{1.0}{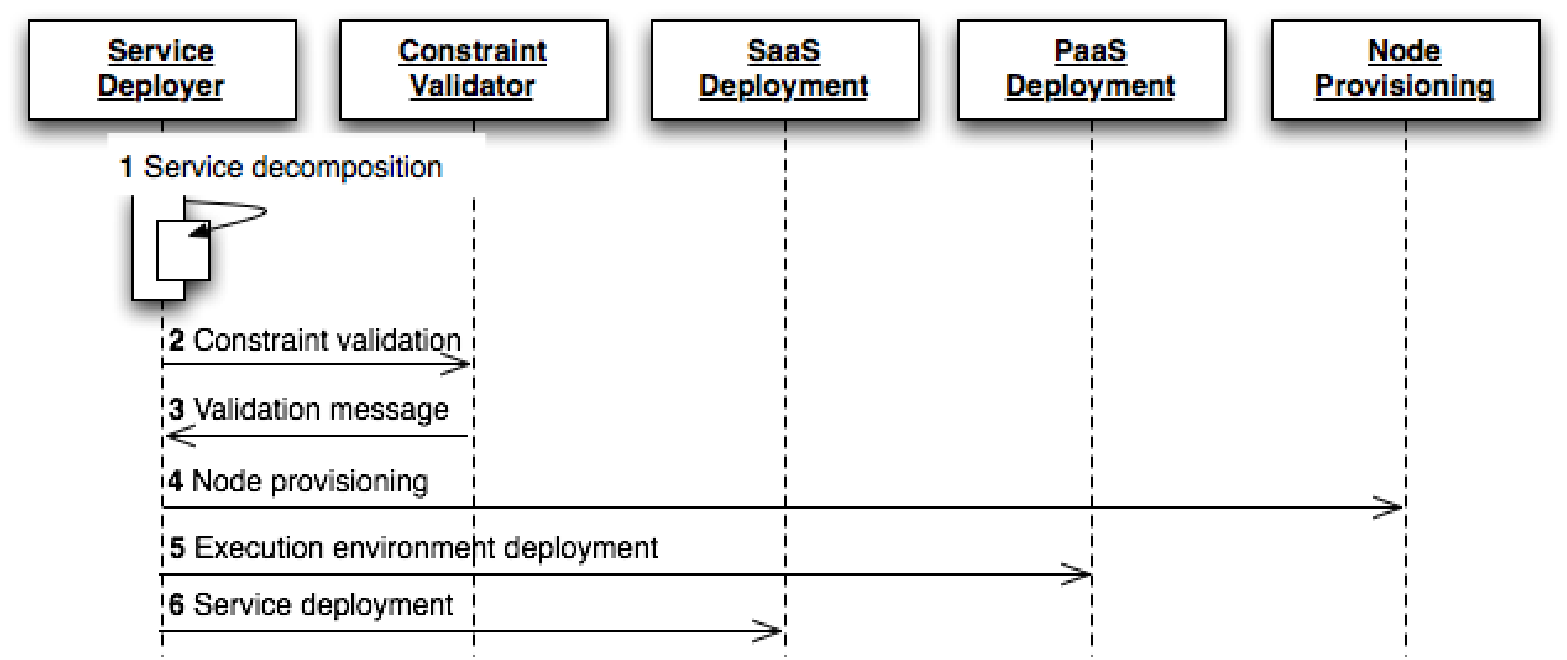}{Application deployment sequence diagram.}
To perform the deployment, the SD component captures the constraints defined and validates them with the \emph{Constraint Validator} component. Once the validation is done, the SD component deploys a whole application this corresponds to sequences 4 to 6 in Fig.~\ref{fig:sequence-diagram} with the support of the SaaS Deployment, PaaS Deployment, and Node Provisioning components.

This component deploys the contribution package in three steps:
\begin{enumerate}
  \item Validates the contribution package by checking if the contribution package contains at least one ZIP file and one composite file.
  \item Uses the constraint validator to validate the SCA properties defined in the composite file.
  \item Matches each constraint or elasticity rule defined in the composite file and invokes the corresponding execution operation: \textit{Node provisioning}, \textit{PaaS deployment}, \textit{SaaS deployment}.
\end{enumerate}
\vspace{-0.7cm}

\subsection{Elasticity specification}
\vspace{-0.2cm}
In this section, we will describe how the soCloud architecture automates specific elasticity rules associated with soCloud applications.

soCloud manages elasticity at IaaS and PaaS levels in the same manner. In fact, the elasticity management in soCloud is not focused on any cloud layer (IaaS or PaaS) specific resources, instead it refers to resources through abstractions provided by the NP component, that offers an uniform way to manage resources from both IaaS and PaaS. soCloud provides the capacity to scale the resources allocated for the application as needed. For example, soCloud can add more nodes if it detects a degradation on the application performance. On the other hand, if the resources are underused, resizing is necessary. This feature is managed as a feedback control loop by the soCloud platform. However, for specific cases, the developer should be able to define automatic elasticity rules associated to its application. These rules are defined inside the application architecture and supervised by the soCloud platform. Each rule is composed of a condition or a set of conditions to be monitored. Those specific elasticity rules are also managed by the soCloud feedback control loop. In order to achieve elasticity, we need to keep track of the frequency of requests to resources hosted and applications deployed on them. Thus, we use a \textit{proactive scheme} that relies on the current workload arrival rate to detect overload conditions. We measure the incoming workload rate by monitoring the number of user connections being opened in the load balancer component. To maintain hit statistics for frequently-accessed applications, we dynamically compute an exponential weighted moving average of request inter-arrival times, along the same lines as TCP computes its estimated round-trip time~\cite{karn1987improving}. Specially, we compute an average of the inter-arrival time using the following formula:

\begin{equation}
f(t) = \left( 1 - \alpha  \right) \ast f(t-1) + \alpha \ast \left( \delta t(t) - \delta t \left( t - 1 \right) \right)  
\label{eq:elast}
\end{equation}

The arrival time of every hit is represented by \( \delta t(t) \). The constant \( \alpha \) is a smoothing factor that puts more weight on recent samples than on old samples. We have used a value of \( \alpha \) = 0.125, which is recommended for TCP\footnote{http://tools.ietf.org/html/rfc2988}.

To detect overloads and underloads in the soCloud platform, we use a \textit{threshold-based} scheme to trigger dynamic allocation. Let us note that the calculation of a \textit{threshold} scheme based on equation~\ref{eq:elast} varies from one deployed application to another.

\vspace{-0.5cm}
\begin{figure*}[htbp]
\begin{center}
\centering
\includegraphics[height=0.7\textwidth, width=1.0\linewidth]{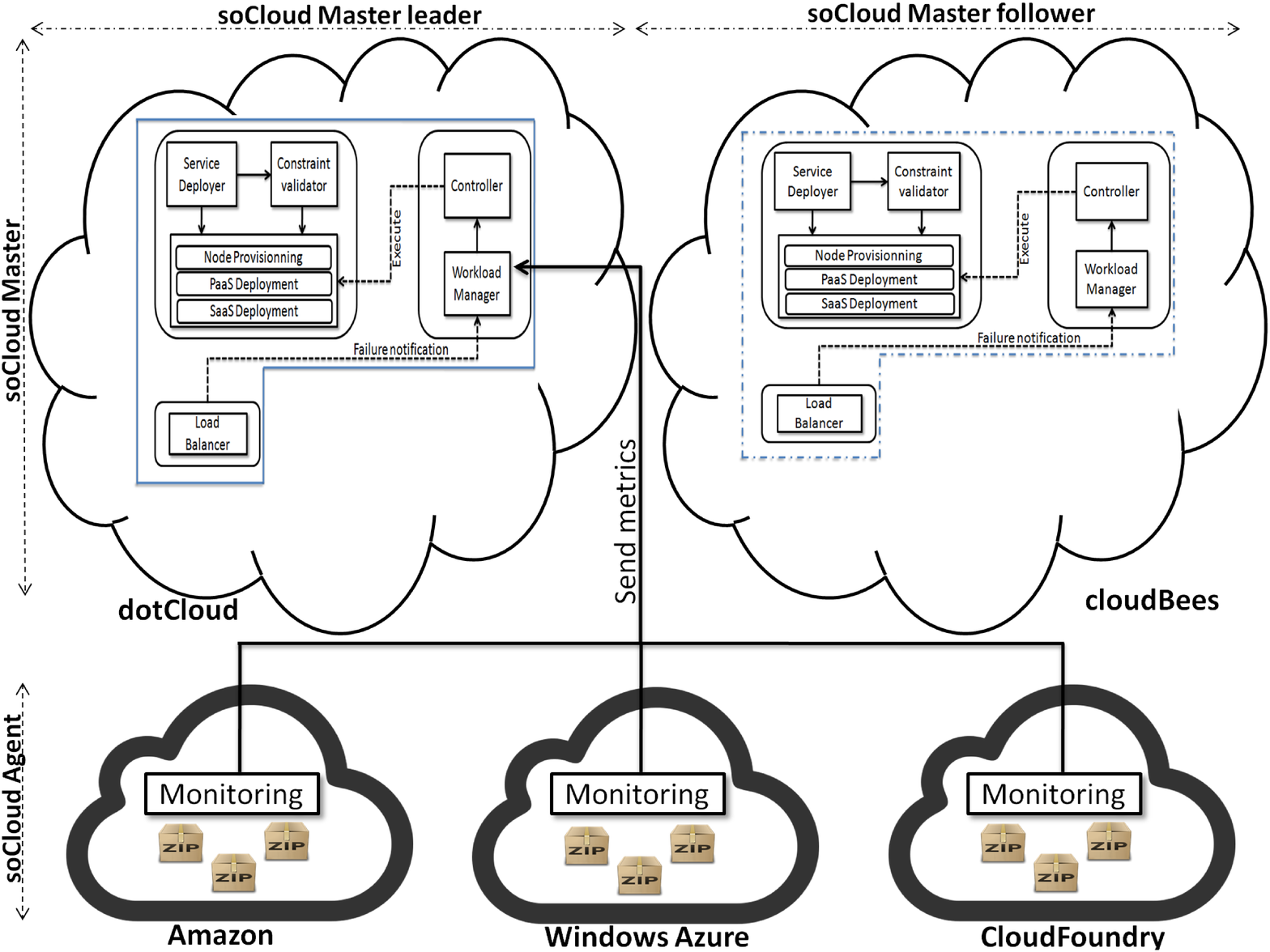}
\caption{Conceptual view of a soCloud deployment.}
\label{fig:mcp-deploy}
\end{center}
\end{figure*}
\vspace{-0.8cm}

\vspace{-0.5cm}
\subsection{soCloud deployment}
\vspace{-0.2cm}
In this section we describe how the soCloud architecture is deployed on a concrete multi-cloud environment.

The soCloud PaaS provides high availability by replicating itself on different clouds as shown in Fig.~\ref{fig:mcp-deploy}. We assume in our implementation that cloud services can fail, and such fault service may later recover. The system administrator has the possibility to define the deployment policy by specifying the number of replications. For instance in Fig.~\ref{fig:mcp-deploy}, there is one replication of the soCloud master. Then, the deployment is done in three steps. In the first step, the soCloud \emph{master} is deployed in dotCloud. In the second step, the soCloud \emph{master} (deployed in dotCloud) dynamically deploys another soCloud \emph{master} in CloudBees. Automatically, the first soCloud \emph{master} becomes leader and the second one the follower. The soCloud \emph{master} leader is active, while the soCloud \emph{master} follower is passive. By active, we mean the soCloud \emph{master} processes the operations in the system. By passive, we refer to the standby soCloud \emph{master} used as replication. At this stage, only the soCloud \emph{master} and its replication are deployed. Finally, the soCloud \emph{master} leader will provision a new cloud node on which it deploys both the execution environment (FraSCAti) and a soCloud \emph{agent}. When the soCloud \emph{agent} is deployed, it uses a service discovery mechanism to find which \emph{Workload Manager} component the information collected should be sent. Periodically, the service discovery checks if the \emph{Workload Manager} component is reachable in order to update the services table when failure occurs on the target soCloud \emph{master}. The soCloud PaaS service discovery mechanism is implemented using Google Fusion Table~\cite{gonzalez2010google} and FraSCAti dynamical multiple reference binding. We use \emph{Google Fusion Table} to persist the state of the active master and with the \emph{dynamical multiple reference} provided by FraSCAti we add on the fly a reference to the new component. By state we mean the operational state of the soCloud \emph{master}. soCloud provides a capability for reliability using sources of state that are external to soCloud itself. Typically, this is done with \emph{Google Fusion Table}. The soCloud platform provides a mechanism called ``health checking'' by which a component notifies its health. This mechanism is implemented as an XML push mechanism which tests if a component is reachable. Both the soCloud \emph{master} and \emph{agent} need the execution environment (FraSCAti) to be running. However, when the system grows (the number of applications or load increase), the third step is repeated.

\subsection{Fail-overs}
\vspace{-0.2cm}
In this section we describe how the soCloud PaaS ensures the high availability at two levels: \textbf{soCloud level}, and \textbf{application level}.
\paragraph{\textbf{soCloud level}}
The active soCloud \emph{master} is called the leader and the passive soCloud \emph{master} is called the follower. The process of electing a leader allows the system to indicate which soCloud \emph{master} will have the decision of execution. The soCloud \emph{master} leader and follower are synchronized such that when the leader fails, automatically the leader election is organized to elect a new leader. Specifically, we use \emph{Wait-Free Synchronization} that is appropriate in fault tolerant and real-time applications~\cite{garg2005concurrent}. In the case the system administrator has been defined only one replication of the soCloud \emph{master}, the soCloud \emph{master} follower is automatically elected. Otherwise, the election is organized between the soCloud \emph{master} followers. The leader election is organized and supervised by the \emph{controller} component. We assume that each component has a \textit{reachable latency}. The \textit{reachable latency} is obtained by making a ping from one component to another. Ping refers to the ability to have a live component connection. Our leader election algorithm is simple. This algorithm ensures that the component with minimum {reachable latency} gets elected as the leader. However, the soCloud platform is not restricted to this algorithm, the system administrator has the possibility to define another one (e.g., Chang-Roberts algorithm~\cite{chang1979improved}, Malpini algorithm~\cite{malpani2000leader}), according to her requirement. Elections are held between two entities that have the same function (e.g., two monitoring components, two workload manager components, etc.). Then, the \emph{controller} component organizes an election in order to compare the \textit{reachable latency}. By using this strategy, all the components of the soCloud \emph{master} leader are in the same cloud and the follower components in other. When the soCloud \emph{master} follower fails, automatically the soCloud master leader deploys a new soCloud \emph{master} follower. 
\vspace{-0.3cm}
\paragraph{\textbf{Application level}}
Same to soCloud replication, the developer has the possibility to define the number of instances which will be deployed for its application. Each application deployed with soCloud is replicated in different clouds. The fail-overs mechanism is achieved by the \emph{LB} component. When failure occurs with one instance of the application, the ~\emph{Controller} component takes the decision to instantiate a new one.

Overall, the fail-overs automation in the soCloud platform enables our system to recover quickly from most outages. In addition, we also monitor our system for any variety of error conditions. With the two levels of availability, the soCloud PaaS addresses the high availability challenge presented in Section~\ref{sec:chal}.4.

\vspace{-0.5cm}
\subsection{Recovery}
\vspace{-0.2cm}
In this section we describe the method used by the soCloud PaaS for fault tolerance, i.e., \emph{check-pointing}\footnote{A checkpoint is a snapshot of the state of a process, saved on nonvolatile storage to survive process failures ~\cite{wang1997consistent}.}. 
A checkpoint can be local to a process or global in the system. With the soCloud PaaS we use a global checkpoint. We use Google Fusion Table~\cite{gonzalez2010google} to record a global state of the system so that in the event of failure the entire system can be rolled back to the global checkpoint and restarted. To record the global state, soCloud uses the coordinated checkpoint method~\cite{bouteiller2003coordinated}. In fact, there are some disadvantages of uncoordinated checkpoint compared with coordinated checkpointing schemes~\cite{garg2005concurrent}. First, for coordinated checkpoint it is sufficient to keep just the most recent global state snapshot in the stable storage. For uncoordinated checkpoints a more complex processing scheme is required. Moreover, in the case of a failure, the recovery method for coordinated checkpoint is simpler.
\vspace{-0.4cm}
\begin{figure*}[htbp]
\begin{center}
\centering
\includegraphics[height=0.6\textwidth, width=1.\linewidth]{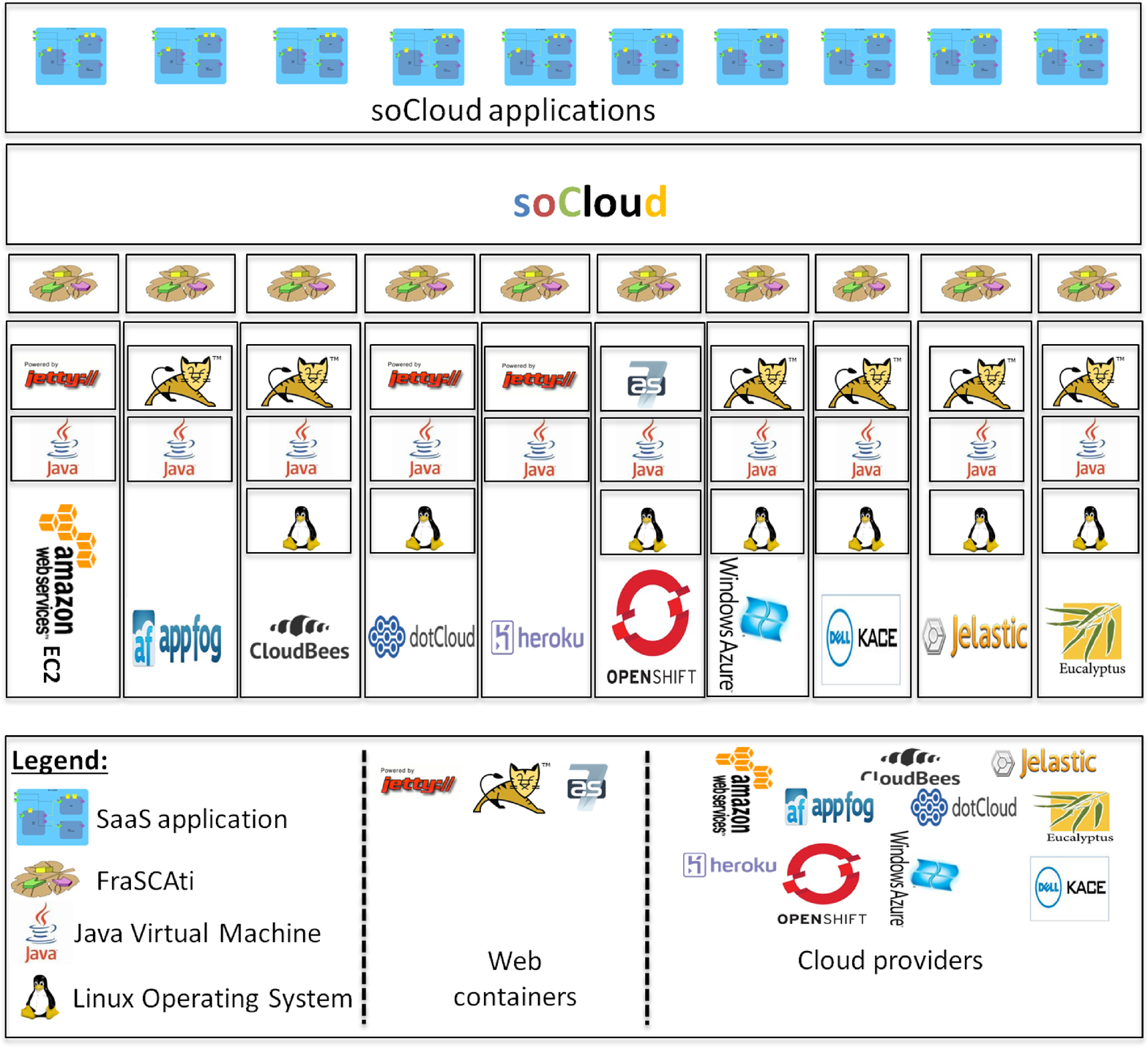}
\caption{soCloud deployment with ten cloud providers.}
\label{fig:soCloud-Integration}
\end{center}
\end{figure*}
\vspace{-0.3cm}

\subsection{Integration with existing IaaS/PaaS}
\vspace{-0.2cm}
We report on the existing cloud environments on which the soCloud platform has been deployed. The soCloud platform extends an experiment that was presented in a previous work~\cite{paraiso2012federated}. The soCloud platform is actually deployed on ten target cloud environments that are publicly accessible on the Internet\footnote{http://socloud.soceda.cloudbees.net}. The deployment is done with IaaS/PaaS providers as illustrated in Fig.~\ref{fig:soCloud-Integration}. With IaaS, resources are provisioned from Windows Azure, DELL KACE, Amazon EC2, and  our Eucalyptus private cloud, we installed a PaaS stack composed of a Linux distribution, a Java Virtual Machine, a web container and FraSCAti. soCloud is also deployed on PaaS such as: CloudBees, OpenShift, dotCloud, Jelastic, Heroku, and Appfog as a WAR file.

\vspace{-0.4cm}
\section{Evaluation}\label{sec:valid}
In this section, we evaluate three key aspects of the soCloud platform: \textit{elasticity}, \textit{high availability} and the overhead introduced by soCloud. Firstly, Section~\ref{subsec:usecase} describes a use case scenario. Then, Section~\ref{subsec:measure} evaluates the reaction of soCloud when faced with flash crowd effects (i.e., elasticity of soCloud). Section~\ref{subsec:high} evaluates the soCloud behavior against failures (i.e., high availability of soCloud). Finally, Section~\ref{sub:over} evaluates the overhead introduced by soCloud.

\vspace{-0.5cm}
\subsection{Use case}\label{subsec:usecase}
\vspace{-0.3cm}
We describe a scenario that can be used in a multiple clouds environment, and explain briefly its requirements.

\vspace{-0.4cm}
\subsubsection{Description}
Let us consider a motivating scenario in which a company built a device called "Fuel optimiser" in charge of reducing the fuel consumed by vehicles (car, boat, tractor, lorry, etc). To improve the quality of their products, they analyse metrics (fuel consumption per km) collected from vehicles. At the end of each trip, the vehicle sensors sent metrics to a company application via REST messages.  The application must face requirements like:
\begin{itemize}
  \item The application must be close to vehicles (geo-diversity).
  \item Unpredictable and unlimited growth of vehicles.
  \item Peaks and unpredictable workloads.
\end{itemize}
To address these challenges, the architecture of this application and the infrastructure need to be flexible, highly available, well performing, reliable and scalable. The application uses a three-tier model; the vehicle sensors are directly connected to the {\bf frontend} tier, the {\bf middle} tier analyses the metrics collected, and the {\bf storage} tier stores the metrics into a database. The application described in this scenario is used for the evaluation of soCloud elasticity, and high availability.

\begin{figure*}[htbp]
\begin{center}
\centering
\includegraphics[height=0.6\textwidth, width=1.0\linewidth]{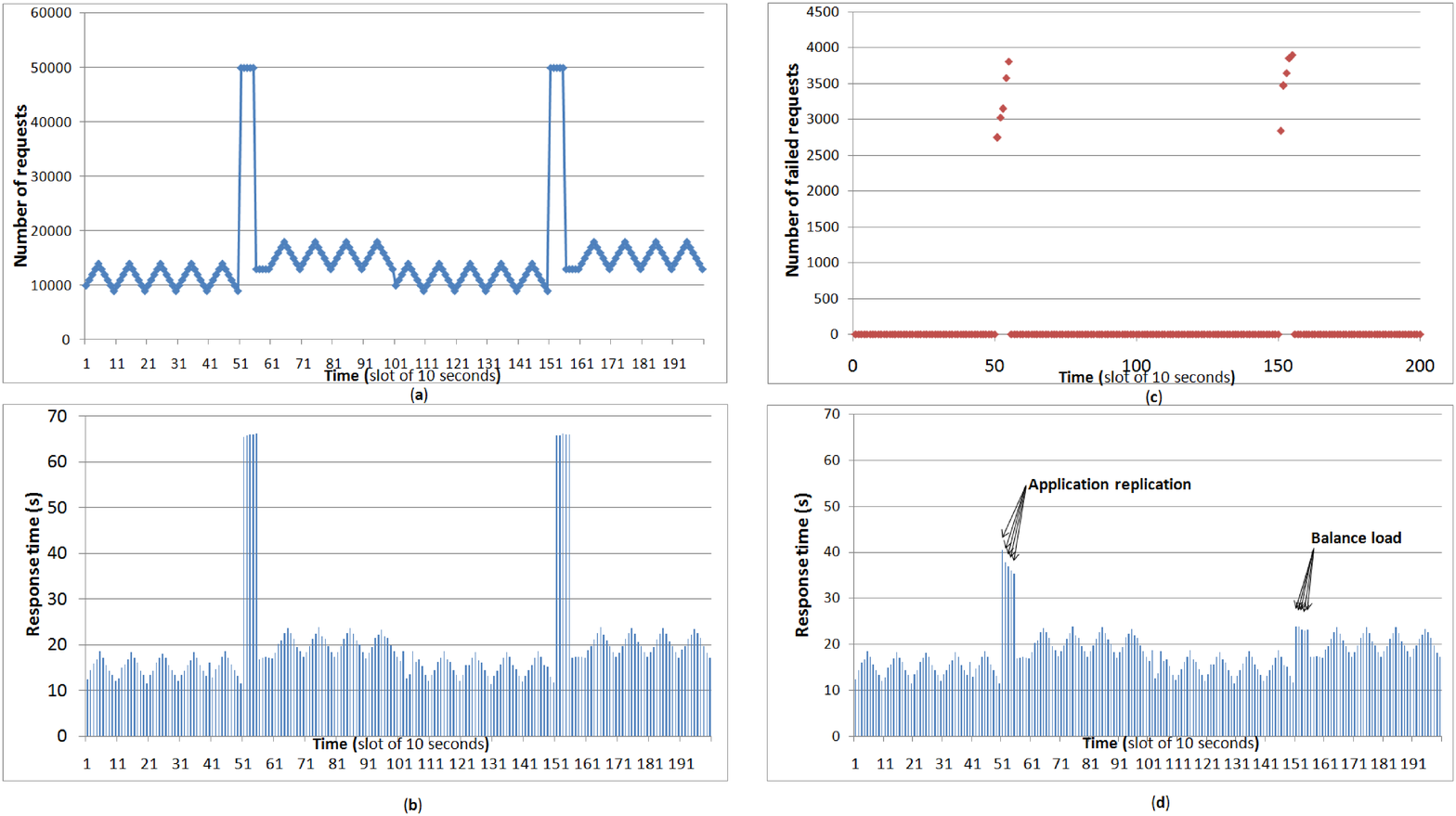}
\caption{\textbf{The series of two flash crowd effects. (a) Effective number of requests during the evolution of the scenario. (b) Response time experienced by clients during the flash crowd effect without soCloud elasticity. (c) Number of requests failed during the two phases of the flash crowd effect. (d) Response time experienced by clients during the flash crowd effect with soCloud elasticity.}}
\label{fig:fc}
\end{center}
\end{figure*}

\vspace{-0.3cm}
\subsection{Measure of the reaction time to flash crowd effects}\label{subsec:measure}
\vspace{-0.2cm}
We have implemented a prototype of the application described in Section 5.1 and deployed it with the soCloud platform. Our soCloud platform is deployed on ten different cloud providers. We have conducted an experimental evaluation of this application in order to assess how the soCloud platform behaves and effectiveness at sustaining flash crowd effects\footnote{The flash crowd effect, also called the slash dot effect, resulte from a sudden increase in request traffic.}. We conducted an analysis of: (a) the application is deployed without the soCloud elasticity mechanism, and (b) the application is deployed with the soCloud elasticity mechanism. 
\vspace{-0.3cm}
\subsubsection{Without soCloud elasticity}
In the first case, we have observed the behavior of this application without elasticity capability under high request load. Each request triggers an operation that consists of analysing metrics collected by a vehicle and stores the results into a database. To that end, we have configured httperf~\cite{Mosberger} to create 50,000 connections, with 10 requests per connection and a number of new connections created per second varying between 10 and 150 ; this corresponds to a total of 3,020,000 requests. Fig.~\ref{fig:fc}\textbf{(a)} shows the number of requests achieved by the application with two phases of a flash crowd effect, and Fig.~\ref{fig:fc}\textbf{(b)} shows the corresponding response time (computed as the number of operations performed). During the two phases of the flash crowd effect, the average response time is 65.90 seconds. Fig.~\ref{fig:fc}\textbf{(a)} and ~\ref{fig:fc}\textbf{(b)} show a mounted sudden load caused by the flash crowd effect. We have noted that the number of requests increases with the response time. Then, Fig.~\ref{fig:fc}\textbf{(c)} shows the number of request errors, and shows the corresponding number of the failed requests. Thus, during the flash crowd effect, 1.13\% of requests have failed, precisely 34,039 requests. These request errors are due to the processing timeout that we have set at 5 seconds for each request. In fact, this timeout means that the lack of any server activity on the TCP connection for this duration will be considered to be an error. 

Overall, when the application becomes saturated, it suffers from performance failures and cause long response delays. We observe that the application can sustain the request rate only up to a certain limit, which directly depends on the number of requests on a time interval.
\vspace{-0.4cm}
\subsubsection{With soCloud elasticity}
In the second case, we have studied the evolution of the response time during the two phases of the flash crowd effect when soCloud elasticity is activated. 

We assume that resource(VM) is preallocated and a soCloud agent is deployed inside. Fig.~\ref{fig:fc}\textbf{(d)} shows the results of the same experiment when using the soCloud elasticity mechanism. We initially observe some contention at the source of application as the response time decreases. During the first phase of the flash crowd effect, the average response time is 37.30 seconds. Indeed, the soCloud platform has detected peak mounted in \emph{300 ms}. After 4 seconds, the soCloud platform replicates the application into another soCloud agent and updates the load balancer table for balancing charge across different instances of the application. This reaction appears clearly in Fig.~\ref{fig:fc}\textbf{(d)}, where the application replication is performed. The soCloud load balancer dispatches the requests among the two instances of the application and the response time remains small despite the high traffic. During the second phase of the flash crowd effect, the application was already deployed, the soCloud platform has detected peak mounted in \emph{300 ms}. As shown in Fig.~\ref{fig:fc}\textbf{(d)}, we do not notice mounted peak during the second phase of the flash crowd effect, and the average response time is 23.38 seconds. The relatively small response time during the second phase of the flash crowd is due to the fact that the soCloud platform has already replicated the application. 

Overall, at the peak of the flash crowd, all the requests are performed with zero failure and relatively acceptable response time, the soCloud platform allows the application to scale more with better quality of service. These results demonstrate that the soCloud platform deals well with elasticity across multiple cloud providers.
\vspace{-0.4cm}
\subsection{soCloud behavior against failures}\label{subsec:high}
\vspace{-0.2cm}
We perform all our evaluation with the application described in the previous sections. To show the behavior in soCloud over time as failures are injected, we deploy soCloud as described in Fig.~\ref{fig:mcp-deploy}. The deployment of soCloud is done on ten clouds. The soCloud master is replicated to tolerate more faults. The leader and follower of the soCloud master are deployed respectively on dotCloud and CloudBees. soCloud agents are deployed on Amazon EC2, Windows Azure, DELL KACE, OpenShift, Jelastic, Heroku, Appfog, and our Eucalyptus private cloud.
\vspace{-0.3cm}
\subsubsection{Deployment time of soCloud}
As described in Section~\ref{sec:over}, the deployment of soCloud consists of the deployment of both a soCloud master and several agents.
\vspace{-3mm}
\paragraph{\textbf{soCloud master}}
The deployment of a soCloud master is done by deploying both leader and follower instances on two different clouds to ensure the high availability. The deployment on each cloud consists of deploying the execution environment (FraSCAti) with the soCloud master. The deployment of a soCloud master takes about \textit{2.1 minutes}.
\vspace{-3mm}
\paragraph{\textbf{soCloud agent}}
We measure the time for the deployment of one soCloud agent. The deployment consists of deploying the execution environment (FraSCAti) with the soCloud agent. The deployment of a soCloud agent takes about \textit{0.9 minute}.

Overall, the average time taken to deploy soCloud with two masters (leader and follower) and one agent is about \textbf{3 minutes}.

\vspace{-0.3cm}
\subsubsection{Failure and recovery of a soCloud master leader}
We assume that soCloud is running and our scenario application is deployed. To simulate a failure, we stop the soCloud master leader in dotCloud. In our observations, soCloud takes about \textit{3.5 minutes} average to recovery and to become operational. soCloud takes less than 200 ms to elect a new leader. The recovery process is performed as follows. First, the soCloud master follower becomes leader after the election and rollbacks the system. Then, a new soCloud master follower is deployed on another cloud. Finally, the soCloud agent discovers automatically the new soCloud master leader. According to ~\cite{iwgcr,cloudoutages}, the average Mean Time To Recovery (MTTR) for public clouds is \textbf{7.5 hours}. As a comparison, the recovery time of soCloud takes only \textbf{3.5 minutes} as shown in Table~\ref{tab:mttr}.

\vspace{-0.9cm}
\begin{center}
\begin{table}[htbp]
\caption{MTTR results}
\begin{tabular}{|c|c|c|c|c|}
\hline
	{\bf } & {\bf MTTR(Hour)} \\
\hline
	soCloud & 0.06 hour\\
\hline
	Public clouds & 7.5 hours\\
\hline
\end{tabular}
\label{tab:mttr}
\end{table}
\end{center}
\vspace{-0.8cm}

\vspace{-0.4cm}
\paragraph{\textbf{Failure and recovery of a soCloud master follower}}
In this case, we simulate the failure of a soCloud master follower in CloudBees, the soCloud master leader detects automatically the failure. The soCloud master leader takes about \textit{1200 ms} to elect and start a new master follower.

\vspace{-0.4cm}
\paragraph{\textbf{Downtime of an application deployed on soCloud}}
The failure of an application deployed with soCloud does not affect its availability. In fact, when a failure occurs, the load balancer takes about \textit{300} ms to detect and switch automatically to another instance of the application deployed.

\vspace{-0.4cm}
\paragraph{\textbf{Downtime of a soCloud agent}}
The failure of a single soCloud agent does not affect the availability of the application deployed on soCloud. The soCloud load balancer allows to redirect the requests to another instance of the application. The soCloud agent deployment and start still take about \textit{0.9 minute}. However, the deployment time of applications that were on the platform depends on the size of these applications.

\subsubsection{The soCloud high availability}
Let us consider the availability equation below~\cite{marcus2003blueprints,torell2004mean}:

\begin{equation}
{Availability} = \frac{MTBF}{MTBF + MTTR}
\label{eq:avail}
\end{equation}

As Equation~\ref{eq:avail} shows, the longer the MTTR is, the worse off a system is. The formula illustrates how both Mean Time Between Failure (MTBF) and MTTR impact the overall availability of a system. As MTTR goes up, availability goes down. To compare the availability of soCloud and public clouds, we must estimate the same MTBF. Then, in a year we assume that the MTBF is 8760 hours. The availability is calculated in Table~\ref{tab:avail}.

\vspace{-.5cm}
\newcolumntype{C}{>{\centering\arraybackslash}X}
\begin{center}
\begin{table}[htbp]
\caption{Availability comparison}
\begin{tabular}{|c|c|c|c|c|}
\hline
	{\bf } & {\bf Availability} \\
\hline
	soCloud &  $\frac{8760}{8760 + 0.06}$ = 99.999\% \\
\hline
	Public clouds & $\frac{8760}{8760 + 7.5}$ = 99.914\% \\
\hline
\end{tabular}
\label{tab:avail}
\end{table}
\end{center}
\vspace{-1cm}

Overall, as shown in Table~\ref{tab:avail}, the availability of public clouds is 99.914\%. As a comparison, the soCloud availability is 99.999\%. This result is close from the expected reliability of mission critical systems (c.f. Section~\ref{sec:chal}.4). The soCloud platform increases high availability. This result demonstrates that soCloud ensures well high availability across multiple clouds.
\vspace{-0.4cm}
\subsection{Overhead introduced by soCloud}\label{sub:over}
\vspace{-0.2cm}
In order to analyse the overhead introduced by the soCloud platform, we have deployed our use case application directly on CloudBees and through the soCloud platform. We have packaged two different archive files. The first archive file is a WAR file, its size is $ 50.7 $ Mb. This file contains the application and the execution environment FraSCAti. The second archive file is a Zip file (an SCA contribution), its size is $ 2.1$ Mb. The second file contains only the application. The WAR and Zip files are deployed respectively on CloudBees and soCloud. The deployment of the WAR and Zip files is performed ten times. Table~\ref{tab:dep_time} reports the average deployment time of each file.

\vspace{-0.5cm}
\begin{center}
\begin{table}[htp]
\caption{Deployment time of the Zip and WAR files}
\resizebox{0.7\textwidth}{!} {
\begin{tabular}{|c|c|c|c|}
\hline
	 {\bf Implementation}  & {\bf File size} & {\bf Avg. deploy. time} \\
\hline
	Zip File (Application)  & 2.1 Mb & $5 301.5$ ms\\ 
\hline 
	WAR File (Application + FraSCAti) & 50.7 Mb & $80 830.8$ ms \\
\hline
\end{tabular}
}
\label{tab:dep_time}
\end{table}
\end{center}
\vspace{-1cm}

As noticed, the deployment time of the application directly on CloudBees is greater than the deployment time on soCloud. This is explained by the size of the WAR file which is greater than the Zip file. In fact, uploading a small file in the network is faster than a big file. When deployed the Zip file on the soCloud platform, the execution environment is already deployed and started. This is not the case of the WAR file which contains the FraSCAti execution environment that will be installed and instantiated on the CloudBees PaaS before deploying the application on it.

To evaluate the overhead introduced by the soCloud platform, $10,000$ requests were generated and sent with the Htt\-perf tool. We evaluate two implementations of this scenario: i) the application without soCloud, and ii) with soCloud. The scenario was executed ten times on each of the two implementations. Table~\ref{tab:over} presents the results of the average execution time for each implementation, as well as the mean overhead introduced by the soCloud platform.

\vspace{-0.5cm}
\begin{center}
\begin{table}[htp]
\caption{Execution time and Overhead}
\resizebox{0.8\textwidth}{!} {
\begin{tabular}{|c|c|c|}
\hline
	 {\bf Implementation}     & {\bf Avg. exec. time} & {\bf soCloud overhead} \\ 
\hline
	(Application + FraSCAti)            & $10.85$ sec & -         \\ 
\hline 
	(Application + FraSCAti + soCloud) & $11.10$ sec & $2.3\%$ \\ 
\hline
\end{tabular}
}
\label{tab:over}
\end{table}
\end{center}
\vspace{-1cm}

From the results presented in Table~\ref{tab:over}, we can notice that the overhead introduced by the soCloud platform is $2.3\%$. This overhead is generated by the soCloud monitoring and the Load Balancing components. The overhead of the monitoring component is due to the information collected for the elasticity. 

Overall, the abstraction provided by the soCloud platform is not free, because it introduces an overhead of $2.3\%$. However, the benefits provided by the soCloud platform in multi-cloud environment outweigh the difference in the execution time.

\vspace{-0.5cm}
\section{Related work}\label{sec:related}
Related to the Inter-Cloud Architectural taxonomy presented in~\cite{grozev2012inter}, soCloud can be classified into the Multi-Cloud service category. This section presents some of the related work to multi-cloud computing challenges discussed in Section~\ref{sec:chal}: \textit{portability}, \textit{provisioning}, \textit{elasticity}, and \textit{high availability} across multiple clouds.

\vspace{-3mm}
\paragraph{\textbf{Multi-cloud portability}}
Portability approaches can be classified into three categories\-~\cite{oberle2010etsi}: \textit{functional portability}, \textit{data portability} and \textit{service enhancement}. The authors\cite{petcu2012portable} of mOSAIC deal with \textit{service enhancement} portability at IaaS and PaaS levels. mOSAIC provides a component-based programming model with asynchronous communication. However, mOSAIC APIs are not standardized and are complex to put at work in practice. Our soCloud solution deals with \textit{service enhancement} portability with an API that runs on existing PaaS and IaaS. soCloud supports both synchronous and asynchronous communications offered by the SCA standard. Moreover, SCA defines an easy way to use portable API. 
The Cloud4SOA~\cite{dandria2012cloud4soa} project deals with the portability between PaaS using a semantic approach. soCloud intends to provide portability using an API based on the SCA standard.

\vspace{-3mm}
\paragraph{\textbf{Multi-cloud provisioning}}
A great deal of research on dynamic resource allocation for physical and virtual machines and clusters of virtual machines~\cite{anedda2010suspending} exists. The work of dynamic provisioning of resource in cloud computing may be classified into two categories. Authors in~\cite{mietzner2008towards} have addressed the problem of provisioning resources at the granularity of VMs. Other authors in~\cite{chase2001managing} have considered the provisioning of resources at a finer granularity of resources. In our work, we consider provisioning at both VM and finer granularity of resources. 
 
The authors in~\cite{foster2006virtual} have addressed the problem of deploying a cluster of virtual machines with given resource configurations across a set of physical machines. While~\cite{czajkowski2005resource} defines a Java API permitting developers to monitor and manage a cluster of Java VMs and to define resource allocation policies for such clusters. Unlike~\cite{foster2006virtual,czajkowski2005resource}, soCloud uses both an application-centric and virtual machine approaches. Using knowledge on application workload and performance goals combined with server usage, soCloud utilizes a more versatile set of automation mechanisms.

\vspace{-3mm}
\paragraph{\textbf{Multi-cloud elasticity}}
Managing elasticity across multiple cloud providers is a challenging issue. However, although managed elasticity through multiple clouds would benefit when outages occur, few solutions are supporting it. For instance, in~\cite{buyya2010intercloud}, the authors present a federated cloud infrastructure approach to provide elasticity for applications, however, they do not take into account elasticity management when outages occur. Another approach was proposed by~\cite{vaquero2011dynamically}, which managed the elasticity with both a controller and a load balancer. However, their solution does not address the management of elasticity through multiple cloud providers. The authors in~\cite{marshall2010elastic} propose a resource manager to manage application elasticity. However, their approach is specific for a single cloud provider.

\vspace{-3mm}
\paragraph{\textbf{Multi-cloud high availability}}
Cloud providers such as Amazon EC2, Windows Azure, Jelastic already provide a load balancer service with a single cloud to distribute load among virtual machines. However, they do not provide load balancing across multiple cloud providers. Different approaches of dynamic load balancing have been proposed in the literature~\cite{cardellini1999dynamic,harchol1997exploiting}, however, they do not provide a mechanism to scale the load balancers themselves. The authors in~\cite{qian2007agility} have explored the agility way to quickly reassign resources. However, their approach does not take into account a multi cloud environment.
Most existing membership protocols~\cite{birman1994reliable} employ a consensus algorithm to achieve agreement on the membership. Achieving consensus in an asynchronous distributed system is impossible without the use of timeouts to bound the time within which an action must take place. Even with the use of timeouts, achieving consensus can be relatively costly in the number of messages transmitted, and in the delays incurred. To avoid such costs, soCloud uses a novel Leader Determined Membership Protocol that does not involve the use of a consensus algorithm.

\vspace{-0.5cm}
\section{Discussion and limitations}\label{sec:discus}
soCloud is a PaaS to aggregate multiple clouds. Throughout the article, we have essentially discussed the advantages of soCloud. On the one hand, soCloud may miss some features that are provided by the underlying clouds used. In other words, soCloud may not exploit the specific features (i.e., elasticity rules, provisioning properties, replication trigger) that is not provided by it. On the other hand it may be the case that some developers or companies may not like to use an SCA-based approach. Indeed, the soCloud adoption can become therefore an issue. One approach for soCloud to address these concerns is to use a wrapper that enable transparent access to cloud provider features. As an SCA-based approach, soCloud offers a solution to deploy and execute service oriented applications. It would be useful in future work for soCloud to overcome the constraint of supporting only SCA-based applications. When deployed the soCloud platform on other PaaS, the scaling mechanism of these platforms is not used by our platform in order to avoid duplicated mechanism.

The cloud platforms and their features provided, especially at the PaaS level are evolving dynamically. However, in general the problem of maintaining the mappings to various cloud providers and managing this evolution to keep up with recent features of our supported clouds are a concern. A common way to address these issues is by wrapping them as soCloud features. However, the use of the future standard for Cloud computing is still the best approach.

soCloud provides an abstraction to hide the heterogeneity and the complexity of the underlying clouds. The solution provided by soCloud can introduce an additional cost (i.e., in term of performance, footprint) to existing IaaS/PaaS environments. However, soCloud provides a uniform way to deploy, execute and manage applications in multi-cloud environments. As benefits, the developer focuses on the cloud rather than troubleshooting implementations, exploits multi-cloud portability, has an efficient management of her applications across multi-cloud. In comparison to heterogeneous ways offered by the several IaaS/PaaS solutions, soCloud provides many benefits.

\vspace{-0.5cm}
\section{Conclusion}\label{sec:conclusion}
In this article, we have proposed soCloud a service-oriented component-based PaaS for managing portability, provisioning, elasticity, and high availability across multiple clouds. soCloud is a distributed PaaS that provides a model for building any multi-cloud SaaS applications. This model is based on an extension of the OASIS SCA standard. We surveyed each of the concepts related to express specific elasticity rules, ensure high availability across multiple clouds and pointed out problematics. To address these problems, this article proposes an architecture, and describes the interactions between each component of this architecture. We explain how the components in a soCloud application descriptor can be annotated with elasticity rules, placement constraints, computation constraints. Based on these annotations, deployable contributions can be loaded and deployed in a suitable manner. The article described the approach used by the soCloud platform to ensure high availability. In particular soCloud takes a wait-free approach to the problem of coordinating components in different clouds and uses load balancer to switch from one application instance to another in case of failures. In comparison, the soCloud's availability with public~\cite{iwgcr} cloud availability, we demonstrate that soCloud ensures high availability in minutes instead of hours. We analyse the flash crowd phenomenon on a use case, and demonstrate how the soCloud platform increases the elasticity of the application. This approach is proactive in the case that the content replication is performed when detecting a traffic surge and anticipating a flash crowd. 

As of future work, we plan to continue our research in the following directions.
First, currently, soCloud manages application's components as contribution file in terms of packaging and deployment. The archive that is referred to by \emph{implementation.contribution} may be an artifact within a larger contribution (i.e., an EAR or WAR file inside a larger ZIP file), or archive may itself be a contribution. Indeed, soCloud will manage and deploy all Java EE archive (WAR, EAR).
Second, we will investigate how the concept of aggregated multiple clouds can be used to reduce the resource provisioning cost, while maintaining the Quality of Service (QoS) to customers who use the resources. Third, as many organizations need to move data from one cloud to another we will work on data portability in a multi-cloud environment.

\section{Acknowledgment}
This work is partially funded by the ANR (French National Research Agency) ARPEGE SocEDA project and the EU FP7 PaaSage project.
\bibliographystyle{spmpsci}
\bibliography{paper}


\end{document}